\documentclass[aps,prb,twocolumn,amsmath,amssymb,superscriptaddress,showpacs]{revtex4-1}

\usepackage[utf8]{inputenc}
\usepackage{amsfonts}
\usepackage{amsmath}
\usepackage{amssymb}
\usepackage{amsthm}
\usepackage{fontenc}
\usepackage{graphicx}
\usepackage{xcolor}
\usepackage{textcomp}
\usepackage{epstopdf}
\usepackage{braket}
\usepackage{mathtools}
\usepackage{amsmath}
\usepackage{dcolumn}

\begin{document}
\newcommand{\abs}[1]{\lvert#1\rvert}
\title{Currents and pseudomagnetic fields in strained graphene rings } 

\author{D. Faria}
\email[D. Faria: ]{daiara.faria@gmail.com}
\affiliation{Instituto de F\'{\i}sica, Universidade Federal Fluminense, Niter\'oi,
Av.\ Litor\^anea sn 24210-340, RJ-Brazil}
\affiliation{Department of Physics and Astronomy, Nanoscale and Quantum Phenomena Institute, Ohio University, Athens, Ohio 45701-2979, USA}
\affiliation{Dahlem Center for Complex Quantum Systems and Fachbereich Physik,
Freie Universit\"at Berlin, 14195 Berlin, Germany}
\author{A. Latg\'e}
\affiliation{Instituto de F\'{\i}sica, Universidade Federal Fluminense, Niter\'oi,
Av.\ Litor\^anea sn 24210-340, RJ-Brazil}
\author{S. E. Ulloa}
\affiliation{Department of Physics and Astronomy, Nanoscale and Quantum Phenomena Institute, Ohio University, Athens, Ohio 45701-2979, USA}
\affiliation{Dahlem Center for Complex Quantum Systems and Fachbereich Physik,
Freie Universit\"at Berlin, 14195 Berlin, Germany}
\author{N. Sandler}
\affiliation{Department of Physics and Astronomy, Nanoscale and Quantum Phenomena Institute, Ohio University, Athens, Ohio 45701-2979, USA}
\affiliation{Dahlem Center for Complex Quantum Systems and Fachbereich Physik,
Freie Universit\"at Berlin, 14195 Berlin, Germany}

\date{\today}

\begin{abstract}
We study the effects of strain on the electronic properties and 
persistent current characteristics of a graphene ring using the Dirac representation.  
For a slightly deformed graphene ring flake, one obtains sizable pseudomagnetic (gauge) 
fields that may effectively reduce or enhance locally the applied magnetic flux through the ring. 
Flux-induced persistent currents in a flat ring have full rotational symmetry throughout the structure; 
in contrast, we show that currents in the presence of a {\em circularly symmetric} 
deformation are strongly inhomogeneous, due to the underlying symmetries of graphene.  
This result illustrates the inherent {\em competition} between the `real' magnetic field and the `pseudo' field 
arising from strains, and suggests an alternative way to probe the strength and symmetries of 
pseudomagnetic fields on graphene systems.
\end{abstract}


\maketitle

The appearance of gauge fields in graphene is a beautiful and experimentally accessible
example of a situation
where  concepts of condensed matter and quantum field theory  converge on 
a physical system. \cite{Vozmediano,GuineaRev} Experimental evidence of `bubble' formation on
particular graphene growth processes, \cite{Levy} and controllable routes to
manipulate graphene bubble morphology \cite{Georgiou,drums,Tomori} have motivated numerous works 
addressing different aspects of
these effects.  In particular, the theoretical description of strained
graphene has been developed significantly, exploring how its electronic properties
are affected on curved and strained surfaces. It is known that elastically deformed graphene can
be mapped onto the Dirac formalism in the continuum
limit by including pseudomagnetic fields \cite{KBA2011,Abedpour,Wakker} and Fermi velocity
renormalization, \cite{Gonzalez} giving rise to {\em local} quantities that depend on
strain but do not break time reversal symmetry. 
A recent contribution reported a space-dependent Fermi
velocity leading to interesting experimental consequences. \cite{Fer1} Lattice-corrected strain 
induced vector potentials in graphene have also been discussed within a tight-binding 
scenario, \cite{Pereira, Alex} although these corrections do not contribute to the 
pseudomagnetic field distribution. \cite{Alex,Fer-condmat, Masir, Salvador} 
Interesting possibilities for observing the pseudomagnetic fields arise from breaking time reversal symmetry in the system via an external magnetic field.  This promotes a most interesting interplay, some of which has been explored in the context of the quantum Hall regime. \cite{Roy1,Roy2}

Similar to other confined systems with periodic boundary conditions, magnetic flux-dependent
persistent currents and conductance oscillations are
expected for graphene rings in an Aharonov-Bohm (AB) geometry. \cite{Morpurgo,Cong,Richter}
Several experiments have verified the presence of AB conductance oscillations with different 
visibility for different device geometries. \cite{Russo,Ihn1,Ihn2,Haug,Rahman}  A recent review
of quantum interference in graphene rings discusses open questions in the field. \cite{Trauzettel}
Interestingly, the `infinite mass' confinement \cite{Berry,Beenakker1,Baranger,Peres} that requires null current density across the boundaries results
in persistent currents that are `valley polarized' in the presence of magnetic flux,\cite{Morpurgo} suggesting that 
graphene quantum rings would be an excellent system to analyze 
the effects of induced curvature. 
The main result of the present Rapid Communication is indeed to show that while a flat (unstrained) graphene ring in the AB 
geometry sustains persistent currents with full rotational symmetry, unavoidable strains in typical systems would result 
in inhomogeneous distributions of currents.  In other words, while the strains alone would result in {\em zero} net persistent
current (since time reversal symmetry is not broken by the pseudomagnetic fields), the competition with the AB
flux 
induces spatially inhomogeneous current 
distributions on the system.  This effect can be seen to arise both from a local rescaling of the Fermi velocity 
as well as by the appearance of gauge fields that result from the elastic deformations.  As such, the persistent currents 
originated by the magnetic AB flux acquire a local character that follows the strain fields. 

Moreover, as the corresponding length scales of the current inhomogeneities are given by the strain fields, 
one can imagine using this effect to measure the strain distribution via a scanning magnetometer that would be 
sensitive to the induced currents.  Alternatively, properly designed strain fields would be used to produce desired current patterns. 

{\em Strained graphene}.
Within a tight-binding model, the effects of lattice deformations may be incorporated into the hopping integrals $t_n$ between
nearest-neighbors, \cite{Ando}
so that $t_{n}=t_{0}+\delta t_{n}=t_{0}\left(1-\beta
\epsilon_{ij}\delta_{n}^{i}\delta_{n}^{j}/a^{2}\right)$, where 
$\beta=|\partial \log t_{0}/\partial \log a| \approx 3$ in graphene,  
$\vec{\delta}_n$ are the nearest-neighbor vectors of a given atom at lattice site $n$, and $t_{0}$ and $a$ are the 
nearest-neighbor hopping integral and distance in the unstrained system, respectively. Indices $i$ and $j$ represent directions on the $2$D plane, with repeated index summation convention throughout. 
The strain tensor,
$\epsilon_{ij}=\frac{1}{2}\left(\partial_{j}u_{i}+\partial_{i}u_{j}+\partial_{i}
h\partial_{j}h\right)$, is characterized by 
$u_{i}$  and $h$, the in- and out-of-plane
deformations, respectively. \cite{Landau}

The resulting Hamiltonian in the presence of inhomogeneous strain and external
magnetic field given by $\vec{B}=\vec{\nabla}\times\vec{A}^{ext}$, can be written in a generalized Dirac
form (in the $K$ valley) as
\begin{equation}
\label{Htotal}
H=-iv_{kj}\sigma_{k}\left(\partial_{j}+i\frac{e}{\hbar}
A_{j}^{ext}\right)-iv_{0}\sigma_{j}\Gamma_{j}+v_{0}\frac{e}{\hbar}\sigma_{j}A^{\delta t}_j \,\,\,,
\end{equation}
where $\sigma_{j}$ are Pauli matrices, and the vector potential with trigonal symmetry 
arising from strains is given by \cite{Ando}
\begin{equation}
\label{Ap}
\vec{A}^{\delta t}_i (\vec{r})=\frac{\Phi_{0}}{2\pi}\left(\frac{-\beta}{2a}\right)\left(
\epsilon_{xx}-\epsilon_{yy},
-2\epsilon_{xy}\right)\,\,\,,
\end{equation}
with $\Phi_{0}=h/e$.  The renormalized Fermi
velocity tensor in Eq.\ (\ref{Htotal}) is position-dependent,\cite{Fer1}
\begin{equation}
\label{aij}
v_{ij}(\vec{r})=v_{0}\left(I_{ij}-\frac{\beta}{4}\left(2\epsilon_{ij}+\eta_{
ij}\epsilon_{kk}\right)\right)\,\,\,,
\end{equation}
where $I$ is the $2\times2$ identity matrix and $v_{0}=3at_{0}/2\approx 10^6$m/s (with $\hbar = 1$). 
Inhomogeneous strains also yield an effective geometric vector potential 
\begin{equation}
\label{gamma}
\Gamma_{i}(\vec{r})=-\frac{\beta}{4}\left(
\partial_{j}\epsilon_{ij}+\frac{1}{2}\partial_{i}\epsilon_{jj}\right)\,\,\,.
\end{equation}
At the $K'$ valley, both $v_{ij}(\vec{r})$ and $\Gamma_{i}(\vec{r})$ are the
same, while the vector potential $A^{\delta t}_{j}(\vec{r})$
changes sign, preserving overall time-reversal symmetry of the system in the
absence of $\vec{A}^{ext}$. 

Diagonalization of the Hamiltonian in a ring geometry results in interesting eigenstates and persistent current
patterns for states at and near the Dirac point (charge neutrality point).  
We first summarize the results for a flat graphene ring in a magnetic flux, \cite{Morpurgo}
to provide a suitable framework for the ring with deformation.  

{\em Unstrained graphene ring}.
We consider a `flat' ring threaded by a magnetic flux $\Phi$,
with $\vec{A}^{ext}=(\Phi / 2\pi r)\hat{\theta}$, so that the Hamiltonian is given by 
\begin{equation}
H_{0}=-iv_{0}\left[\Lambda_{1}(\theta)\partial_{r}+\Lambda_{2}(\theta)\frac{1}{r
}\left(\partial_{\theta}+i\frac{\Phi}{\Phi_{0}}\right)\right]\,\,\,,
\label{eq-H0}
\end{equation}
with $\Lambda_{1}(\theta)=\sigma_{x}\cos\theta+\sigma_{y}\sin\theta$ and
$\Lambda_{2}(\theta)=-\sigma_{x}\sin\theta+\sigma_{y}\cos\theta$.
The wave functions for energy $E$ are \cite{Morpurgo}
\begin{equation}
\psi_{\bar{m},s}(r,\theta)=e^{im\theta}\left(\begin{array}{c}
\phi_{\bar{m}}(r) \\
i s e^{i\theta}\phi_{\bar{m}+1} (r)
\end{array}\right)\,\,\,,
\label{psi-eq}
\end{equation}
with
$\phi_{\bar{m}}(r)= A_{\bar{m}} J_{\bar{m}}(kr)+B_{\bar{m}}Y_{\bar{m}}(kr)$, 
where $\bar{m}=m+{\Phi}/{\Phi_{0}}$, 
$m=0,\pm1,\pm2, ...$ is the orbital angular momentum, $J_{\bar{m}}$ and $Y_{\bar{m}}$ are Bessel functions of first and second kind, respectively, $k = |E| /\hbar v_0$, and $s={\rm sgn}(E)$. The upper (lower) component of the spinor corresponds to $\psi_A$ $(\psi_B)$ in Eq.\ (\ref{psi-eq}) for the $K$ valley.

The energy spectrum is obtained from the transcendental equation
that arises after imposing infinite-mass boundary
conditions, \cite{Morpurgo,Beenakker1} given by
$%
{\psi_{B}(r,\theta)}/{\psi_{A}(r,\theta)}
=i \tau \hat{n}\cdot\hat{r} e^{i\tau \theta} 
$, 
where the normal to the boundaries is 
$\hat{n}=\pm\hat{r}$ for the inner
$(-)$ and outer radius $(+)$ of the ring. The $K$ and $K'$ valley eigenstates for a given $m$ have also a total angular momentum $j$, defined for the operator $J_{z}=-i\partial_{\phi}+\sigma_{z}\tau/2$, where $\tau=\pm 1$ identifies the valleys; $K$ and $K'$ states with opposite momentum $j'=-j$ are related by $m'=-(m+1)$.  Notice that the boundary
conditions do not mix the valleys and in fact break the valley degeneracy for nonzero flux. \cite{Morpurgo,Beenakker1} 

\begin{figure}[h!]
\centering
\includegraphics[scale=.55]{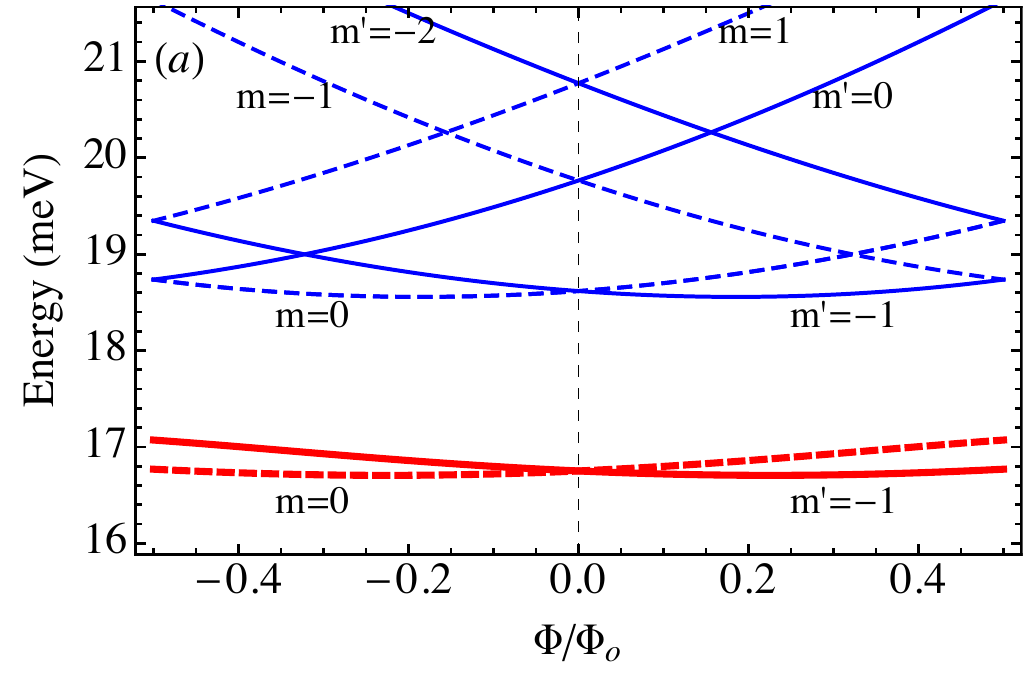}
\includegraphics[scale=.48]{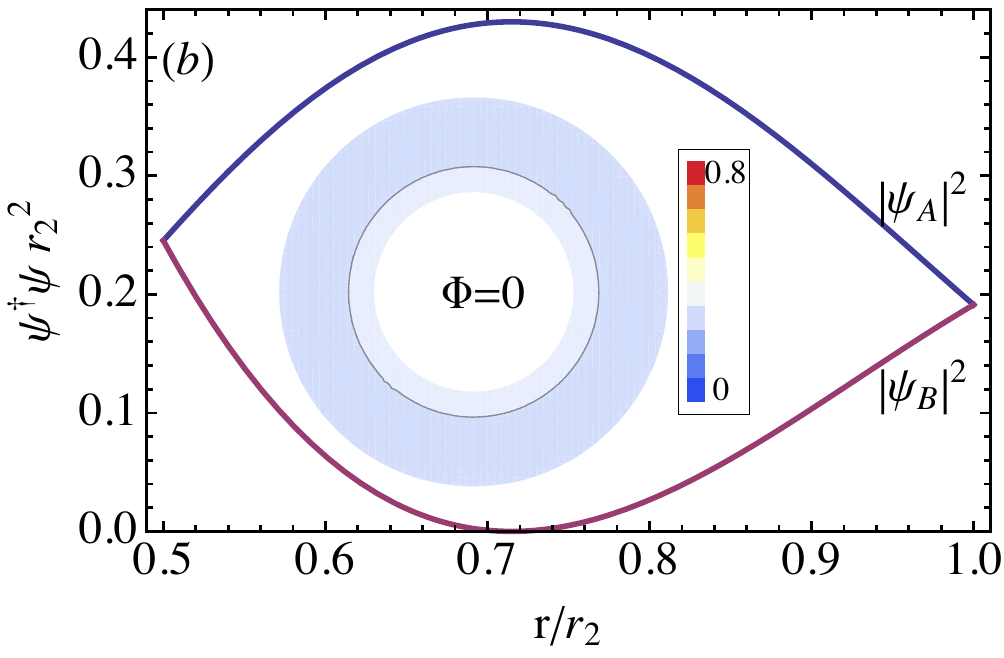}
\includegraphics[scale=.50]{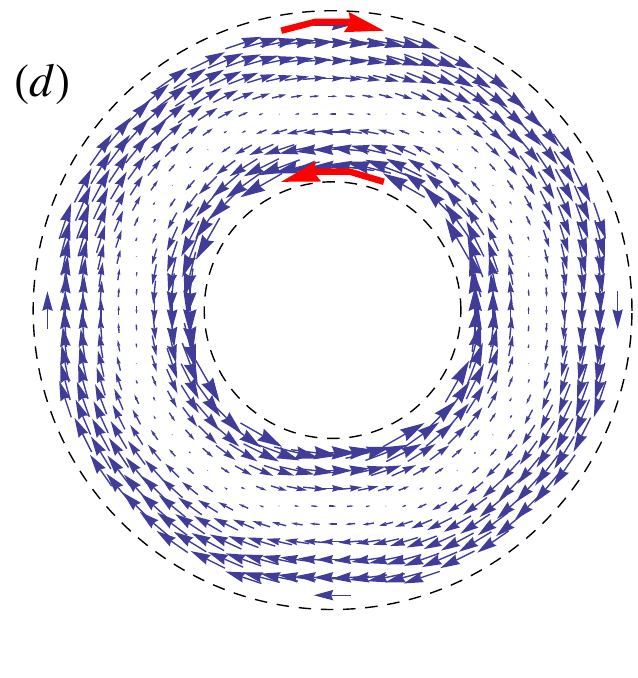}
\includegraphics[scale=.48]{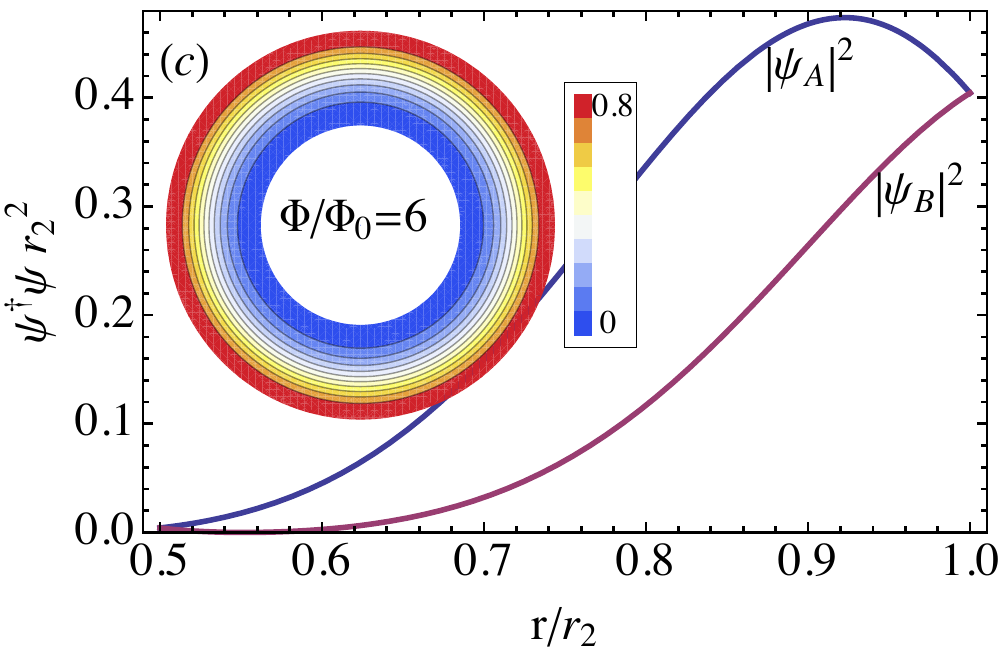}
\includegraphics[scale=.50]{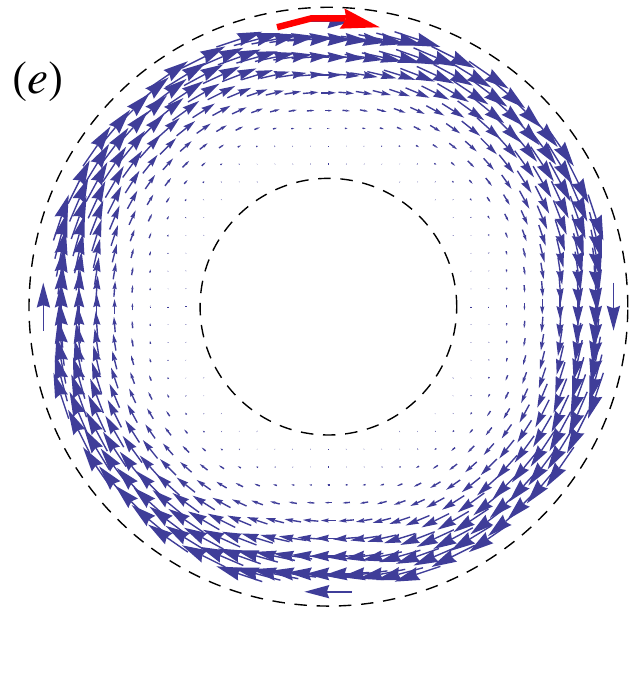}
\caption{(Color online) Unstrained graphene ring. (a) Energy spectra vs\ magnetic flux for ring
with internal and external radii given by $r_1=50$ and $r_2=100$nm, for
different quantum states: $m$ $(m'=-(m+1))$ integer denotes results for $K$ $(K')$
valley given by dashed (continuous) lines. [Thicker (red) lines near $E\approx17 $meV show lower levels for the same ring, but with a 
Gaussian deformation.] 
(b) Flat ring eigenstate with $m=0$ ($K$ valley); electronic
probability distribution along the ring for $\Phi=0$, and (c) $\Phi/\Phi_0= 6$.  
Main panels show the two spinor components separately; insets show total  
distribution, $|\psi_A|^2+|\psi_B|^2$. On right column, corresponding current densities for $K$ valley.  (d) Notice 
counterpropagating edge currents for $\Phi=0$, indicated by red arrows, evolve to a current distribution mostly on the
outer radius for large $\Phi$ in (e).} 
\label{fig1+2} 
\end{figure}

Figure \ref{fig1+2}(a) shows the energy spectrum of a graphene ring vs\ magnetic flux; dashed and continuous 
(blue) lines (upper part of the graph) indicate results for $K$ and $K'$ valleys with $m$ and $m'$ values, respectively. 
Notice the quadratic dependence on $\Phi$ of these levels, which breaks valley degeneracies in general.
The figure also shows the lowest energy levels for a deformed ring (thicker red lines, lower part of graph), to be
discussed later.  

The charge and current densities that satisfy the continuity equation for
unstrained graphene are given by $\rho= \psi^{\dagger}\psi$ and
$J_j= (v_0/\hbar) \psi^{\dagger} \tau \sigma _j \psi$. \cite{Beenakker1}
Typical results for the spinor components $|\Psi_A|^2$ and $|\Psi_B|^2$ for $m=0$ are
depicted in Fig.\ \ref{fig1+2} along the radial direction, for both (b) $\Phi/\Phi_0= 0$, and (c) $6$. As expected, 
increasing flux causes the charge density to be driven to the outer edge of
the ring as the energy of the state increases. (Notice the $m=0$ state is not in general the lowest state
as $\Phi$ increases.) Also shown in Fig.\ \ref{fig1+2}(d) and (e) are the current probability densities,
highlighting the strong dependence on the magnetic flux.  Notice that the current for valley $K$ at zero flux is given by nearly compensated
counter propagating edges, while as the flux increases, the current in the inner edge disappears.  
As we will see below, the interaction with the pseudomagnetic field generated by
the deformation gives rise to an intricate current pattern.

{\em Strained graphene rings}. 
We now consider an out-of-plane 
deformation given by a circularly symmetric Gaussian shape \cite{Fer2} described by 
$h=Ae^{-r^{2}/b^{2}}$.  The strain tensor is then
\begin{equation}
\epsilon=\alpha f(r)\left(\begin{array}{cc}
\cos^{2}\theta &\sin\theta \cos\theta\\
\sin\theta \cos\theta& \sin^{2}\theta
\end{array}\right)\,\,\,,
\end{equation}
where $f(r)=2\left(r^{2}/b^{2}\right)e^{-2r^{2}/b^{2}}$, with
$\alpha=A^{2}/b^{2}$ characterizing the strength of the strain
perturbation.  
The strain is inhomogeneous and, as a consequence, the geometric gauge field 
$\Gamma$ is nonzero and the
renormalized Fermi velocity changes along the ring.
The space-dependent velocity is given by
\begin{equation}
v=v_{0}\left(\frac{-\beta\alpha}{2}\right) f(r)\left[ I 
+ \frac{1}{2}R\left(2\theta\right)
\sigma_z
\right]\,\,\,,
\end{equation}
where $R\left(\theta\right) =  I\cos\theta-i\sigma_{y}\sin\theta$ is the rotation matrix through an 
angle $\theta$ in the counterclockwise direction.  
The resulting Dirac cone becomes elliptical, with radial and angular components
\begin{equation}
v_{r}= v_{0} \left(\frac{-3\beta\alpha}{4}\right) f(r)\,\,\,\, \text{and} \,\,\,\, v_{\theta}= v_{0} \left(\frac{-\beta\alpha}{4}\right) 
f(r)\,\,\,,
\end{equation}
while the gauge fields are
\begin{equation}
\Gamma_{r}= \left(\frac{-\beta\alpha}{2}\right)f(r)\left(\frac{2}{r}-\frac{3r}{b^2}\right)\,\,\,\,\,\,\, \text{and} \,\,\,\,\,\,\,\Gamma_{\theta}=0\,\,\,.
\end{equation}
These changes can be seen as perturbations of the 
Hamiltonian in Eq.\ (\ref{eq-H0}) given by
\begin{equation}
V_{1}=-iv_{0}\left(\frac{-\beta\alpha}{2}\right)
f(r)\left[\Lambda_{1}(\theta)d_{r}+\Lambda_{2}(\theta)\frac{d_{\theta}}{r}\right
]\,\,\,,
\end{equation}
with $d_r=\frac{3}{2}\partial_{r}+\frac{2}{r}-\frac{3r}{b^2}$ and $d_{\theta}=\frac{1}{2}\left(\partial_{\theta}+i\frac{\Phi}{\Phi_{0}}\right)$.
The vector potential perturbation 
$%
V_{2}=v_{0}\left(\frac{-\beta\alpha}{2a}\right)f(r)\Lambda_{1}(-2\theta) 
$ %
is associated with a pseudomagnetic field
$\vec{B}_{\delta t}$, given by
\begin{equation}
\vec{B}_{\delta
t}=\hat{z} \frac{\Phi_{0}}{2\pi}\left(\frac{-\beta\alpha}{2a}\right)f(r)\frac{4r}{b^{2}}\sin\left(3\theta\right)
\,\,\,.
\label{eq-field}
\end{equation}
The eigenvalue problem with $H=H_0 + V_1 + V_2$ can be solved using perturbation theory on the parameter 
$\alpha$ up to second order, keeping sufficient states to achieve full convergence of the results. 

We now present our main results for strained rings considering the Gaussian
perturbation with characteristic system parameters: $A=7$nm
and $b=70$nm, with a relative deformation $\alpha=1\%$, and the 
ring radii used in Fig.\ \ref{fig1+2}(a). 
The two lowest states of the spectrum corrected by the 
Gaussian deformation are shown in Fig.\ \ref{fig1+2}(a) in thick (red) curves near
the bottom of the panel, both 
for solutions near $K$ (dashed lines) and $K'$ (solid) valleys. 
We find that the main correction comes from the $V_2$ perturbation which contains the effects of the strain-induced pseudomagnetic field, and produces energy shifts for the ground state as high as 10\%.

\begin{figure}[h!]
\centering
\includegraphics[scale=0.15]{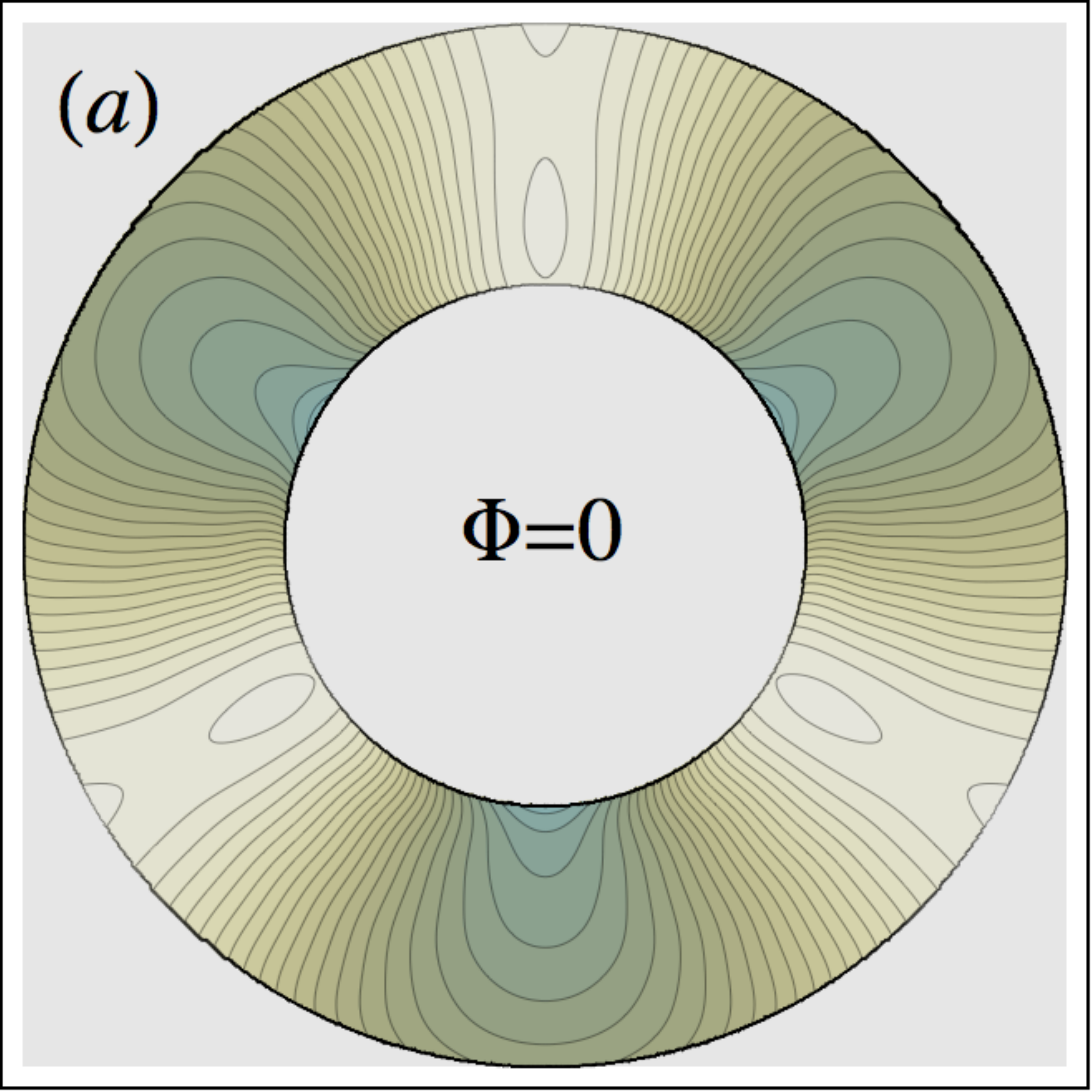}
\includegraphics[scale=0.15]{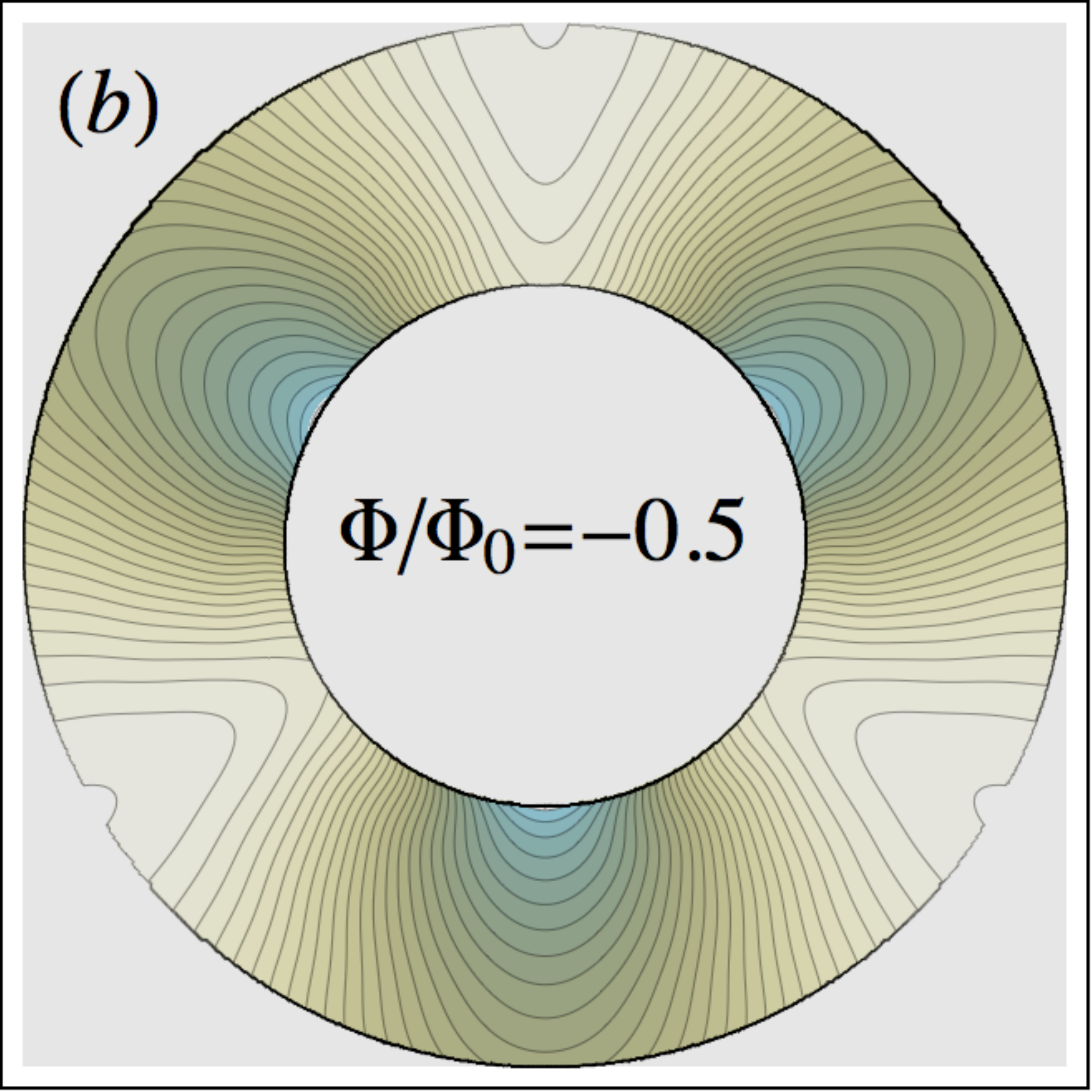}
\includegraphics[scale=.7]{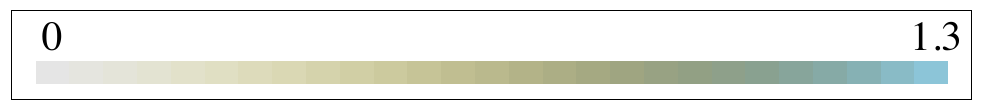}
\includegraphics[scale=0.62]{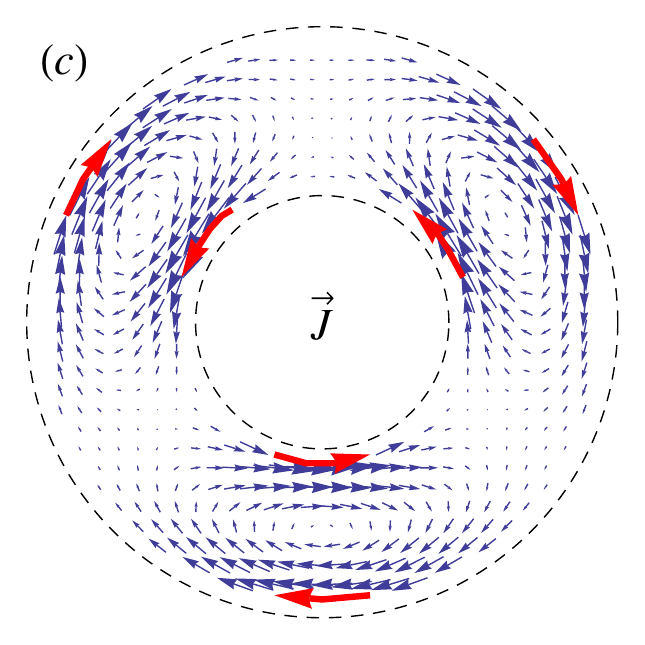}
\includegraphics[scale=0.62]{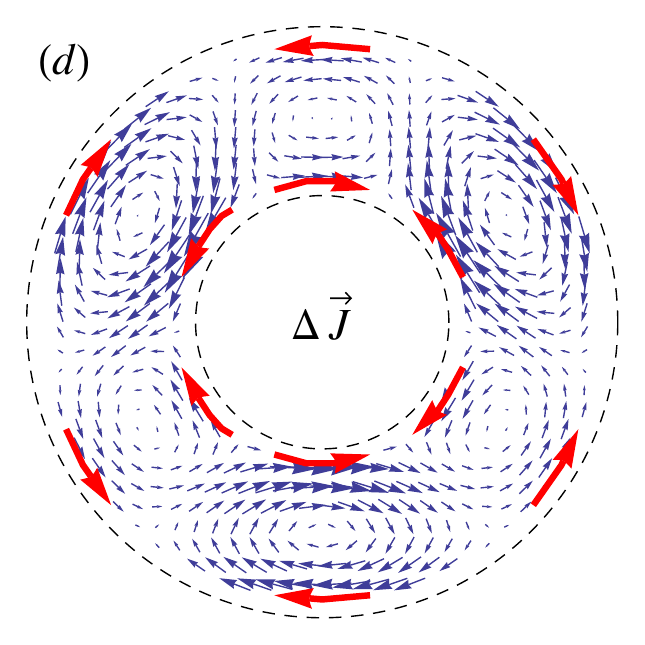}
\caption{(Color online) Deformed graphene ring. Contour plot of the local density 
$|\psi_A|^2+|\psi_B|^2$ for $m=0$ state at (a) $\Phi/\Phi_0=0$, and (b) $\Phi=-0.5$.   
A negative flux, as shown, pushes density towards the inner ring, while a positive
flux would shift weights to the outer radius.  
(c) Current distribution for $\Phi=0$ and $m=0$, for the $K$ valley.
(d) Current variation, $\Delta \vec{J} = \vec{J} - \vec{J}_{\rm flat}$ ($K$-valley), 
where $\vec{J}_{\rm flat}$ is the current for a flat ring--shown in Fig.\ \ref{fig1+2}(d) --can be seen as the net effect produced by the deformation. Notice six vortices with alternating circulation and cores near maxima/minima of the pseudomagnetic field [see Fig.\ \ref{currentx}(a)]. } 
\label{fig-rings}
\end{figure}

\begin{figure}[h!] 
\centering
\includegraphics[scale=0.282]{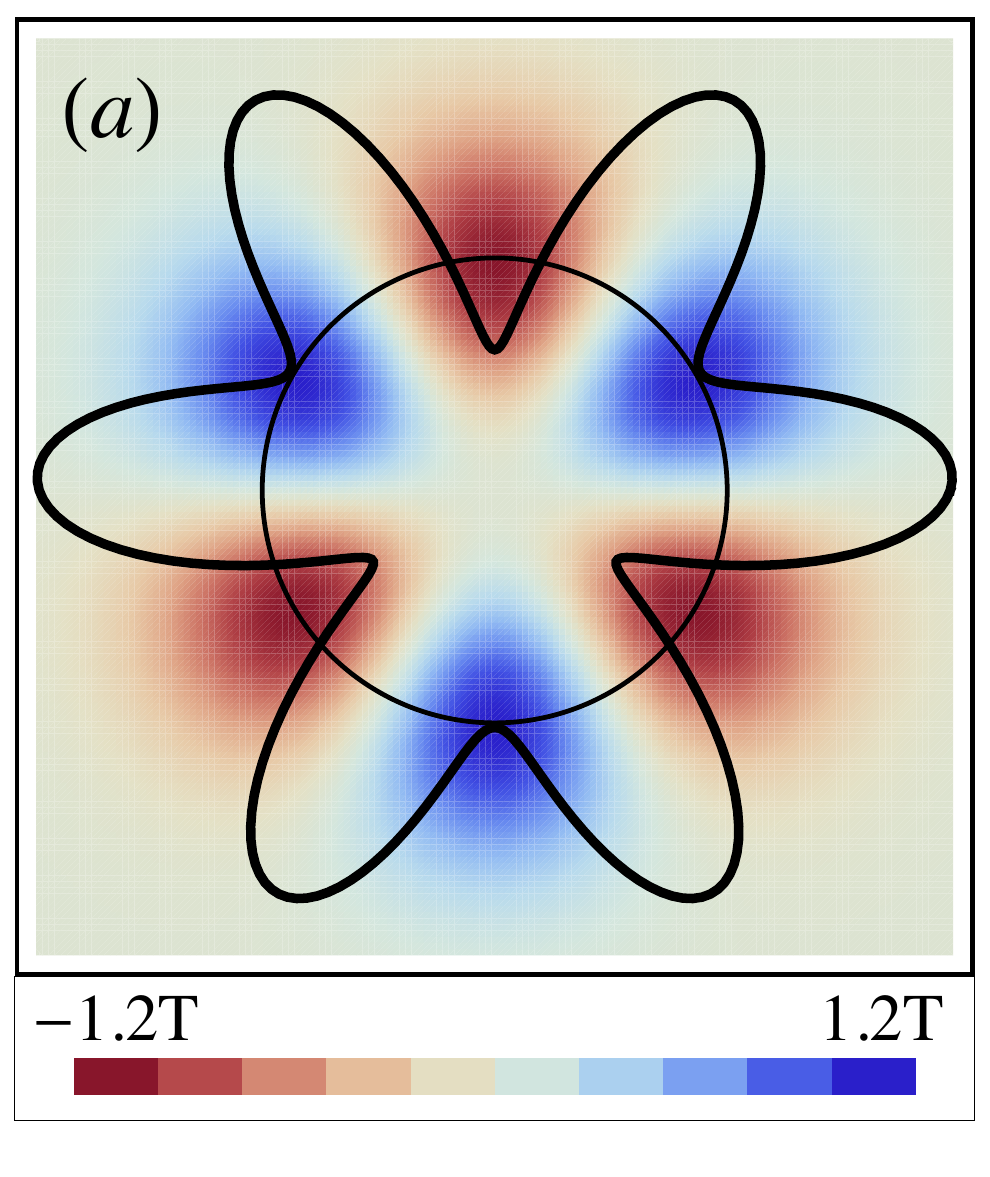}
\includegraphics[scale=0.5]{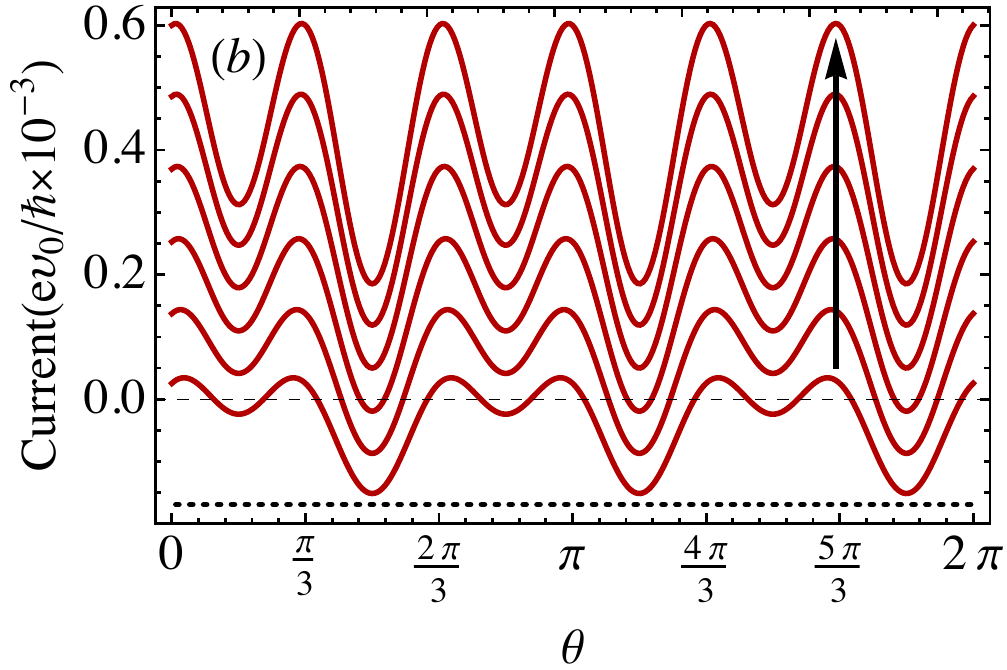}
\includegraphics[scale=0.62]{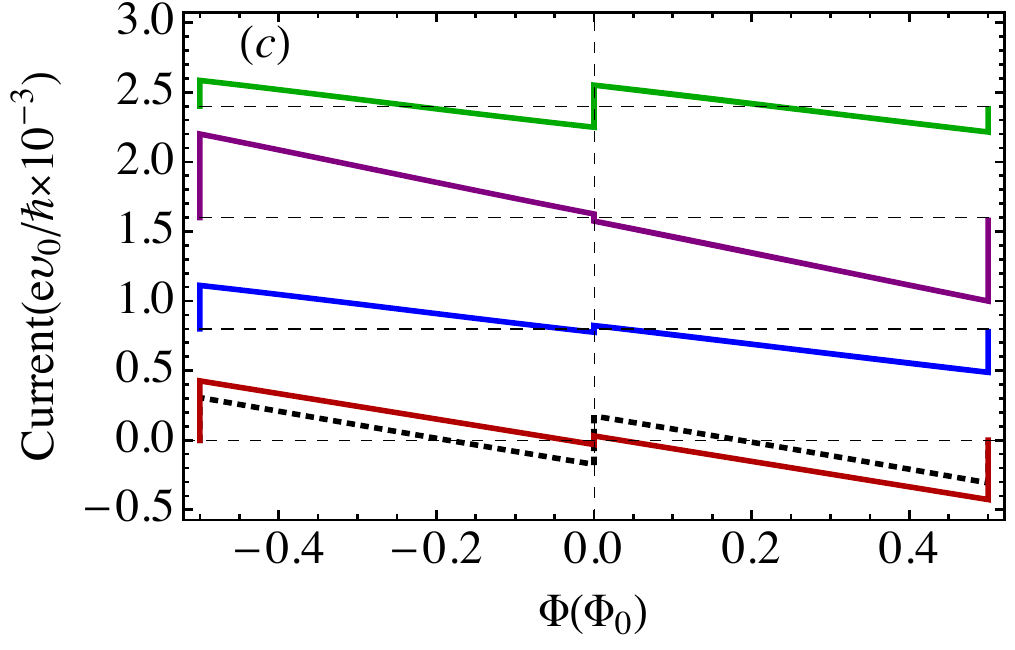}
\includegraphics[scale=0.6]{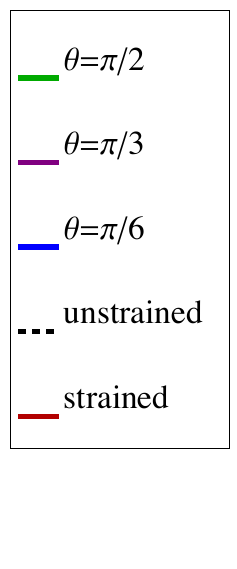}
\caption{(Color online) (a) Map of pseudomagnetic field in Eq.\ (\ref{eq-pseudofield}).  Traces show polar plots of the
current with $\Phi/\Phi_0=-0.5$ for ring without (circular) and with (star) strain.
 (b) Angular profiles of current density through the ring for the lowest $K$ valley state and
different fluxes $(\Delta \Phi/\Phi_0=0.1)$, from $\Phi=0$ to $\Phi/\Phi_0=-0.5$, 
as indicated by the arrow. Dashed flat line near bottom shows current density for unstrained ring at $\Phi=0$. 
Notice mean value of current increases with strain and flux. 
(c) Flux dependence of the current through the ring for {\em the lowest state}: $m=0$, $K$valley from 
$\Phi/\Phi_0=-0.5$ to $0$, and $m=-1$ for $K'$ valley from $\Phi/\Phi_0=0$ to $0.5$, at different angles 
along the ring. Curves for strained rings are shifted up for clarity; dashed lines indicate zero current in each case.
Bottom traces show angle-integrated current for both strained (solid) and unstrained (dotted) rings; notice similar slopes,  
although much smaller discontinuity near $\Phi=0$ with strain. } 
\label{currentx} 
\end{figure}

The pseudomagnetic field produced by the Gaussian deformation in this system
(see Fig.\ \ref{currentx}(a)) has the underlying trigonal symmetry of the graphene lattice.\cite{Wakker,KBA2011}  In this case,
the field amplitudes reach $\simeq 1.2$T, and, as expected, when averaged over the
entire ring, the net pseudofield vanishes.  

When both the external magnetic flux and deformation strains are
considered, the superposition of fields with different symmetry strongly affects the
electronic states and induces inhomogeneities in the probability density distribution.
Typical changes in the spatial pattern of electron density are shown in Figs.\ \ref{fig-rings}(a) and \ref{fig-rings}(b), where zero and finite fluxes are considered. The local density 
is shown by colored projections along the ring.  A finite flux, either positive or negative,
enhances the amplitude modulations of the local probability density seen for zero flux: A negative flux 
shifts the maxima towards the inner radius of the ring, while a positive flux pushes the density  
towards the outer radius, as one would expect from classical Lorentz force considerations.  

The current density $\vec{J}$ over the strained graphene ring is displayed in
Fig.\ \ref{fig-rings}(c), for the lowest state in the $K$ valley.
The current density trends are represented by a set of small (red) arrows, revealing
intricate current configurations. The current density exhibits local maxima with trigonal
symmetry, centered near regions of largest variation in the local density.
The last panel, Fig.\ \ref{fig-rings}(d), shows the probability current variation, $\Delta \vec{J}=\vec{J} - \vec{J}_{\rm flat}$, 
where $\vec{J}_{\rm flat}$ is the persistent current in the flat or unstrained ring. Notice that $\Delta \vec{J}$ exhibits six vortices 
with alternating circulation and cores centered on regions of
extremal values (positive or negative) of the pseudomagnetic field in Eq.\ (\ref{eq-pseudofield}), Fig.\ \ref{currentx}(a).

The spatial modulation of the persistent current in the ring can also be seen from 
the angular profiles shown in Fig.\ \ref{currentx}(b) for
different flux values.  
The current density achieves maximal values along nodal lines of the pseudomagnetic field 
distribution.  Similarly, local minima are related to regions of maximum 
pseudofield amplitude, both in positive and negative directions--see also polar plots in Fig.\ \ref{currentx}(a). As the pseudofield 
does not break time-reversal symmetry, the net current must be zero in the absence of external magnetic flux. 
This indeed happens when considering the contribution of the other valley ($K'$, coming from the state with $m'=-1$), 
which has fully inverted symmetry in $\theta$. 

We analyze the role of the pseudomagnetic field contribution to the total current 
(taking both valleys into account) by looking at different
angular values along the ring.  Figure \ref{currentx}(c)
shows the total current at  $\theta=\pi/6$, $\pi/3$, and $\pi/2$, in comparison with the (angle integrated) current for 
both the unstrained and deformed graphene ring.  All curves present the expected sawtooth behavior with flux. 
However, the 
slope of the curve and the value near zero flux are clearly angle and strain dependent. Notice in particular the 
jump reversal near $\Phi=0$ for $\theta=\pi/6$ and $\pi/3$, associated with the circulation around the vortex
at $\theta \simeq \pi/6$.  
The competition between the external magnetic field and the pseudofield not only results in
inhomogeneous current distributions with vortices, but also in very different total current dependence
with flux $\Phi$.  
Notice that the strain changes in the total persistent current, near $\Phi=0$, are of the same order of magnitude as the current in the unstrained system. As such, the strain decreases the current discontinuity for positive and negative magnetic fluxes. The total current variation dependence on strain is found to be proportional to $\alpha^2$ (not shown).

{\em Conclusions}. 
We have shown that the strain effects arising from a Gaussian `bubble' deformation 
of the graphene ring result in a distribution of pseudomagnetic (gauge) fields
that have trigonal symmetry, in agreement with the underlying symmetries of
graphene.  While the currents induced by these pseudofields would identically vanish, an external magnetic flux makes
possible the observation of the full spatial distribution of currents due to strain. As a result, strain fields change the
nature of the ground state and modify the amount of current present in the device.

Our discussion has focused on the infinite-mass boundary condition. We find also that the 
zig-zag boundary condition, which does not mix valleys either and shows a completely different spectrum, 
yields qualitatively similar results to those presented here. It is clear that an experiment would typically exhibit
a more complex edge.\cite{Zarenia2} However, because the effects are produced by strains fields, there will always be inhomogeneous
current distributions upon the application of a flux.  

Furthermore, other geometric structures with strong strain fields would also produce complex patterns of induced
currents reflecting the pseudomagnetic field distribution. These could  be explored locally by applying a weak magnetic field, to play the role of the external flux,  assuming that
the scanning current measurement device has spatial resolution better
than the characteristic length scales of the gauge field distribution. A scanning magnetometer appears as one of the ideal instruments to reveal these effects.
Although strong sample disorder may increase scan noise, the trigonal symmetry of the strain signal would uniquely identify its source.

{\em Acknowledgments}.
We thank F. de Juan, M. Vozmediano and M. M. Asmar for useful discussions.
This work was partially supported by NSF-PIRE, and NSF and CNPq under the CIAM/MWN program.
AL acknowledges FAPERJ support under grant
E-26/101.522/2010; DF acknowledges support from CNPq (140032/2009-6) and CAPES (2412110) while visiting Ohio University, and from DAAD while at Freie Universit\"at.
We are grateful for the welcoming environment at the Dahlem Center and the support
of the A. von Humboldt Foundation.


\begin{thebibliography}{99}
\bibitem{Vozmediano} M. A. H. Vozmediano, M. I. Katsnelson, and F. Guinea,
Phys. Rep. {\bf 496}, 109 (2010).

\bibitem{GuineaRev} F. Guinea, Solid State Commun.\ {\bf 152}, 1437 (2012). 

\bibitem{Levy}
 N. Levy, S. A. Burke, K. L. Meaker, M. Panlasigui, A. Zettl, F. Guinea, A. H. Castro Neto, and M. F. Crommie, Science {\bf 329}, 544 (2010).

\bibitem{Georgiou} 
T. Georgiou, L. Britnell, P. Blake, R. V. Gorbachev, A. Gholinia, A. K. Geim, C. Casiraghi, and K. S. Novoselov, Appl. Phys. Lett. {\bf 99}, 093103 (2011).

\bibitem{drums} 
N. N. Klimov, S. Jung, S. Zhu, T. Li, C. A. Wright, S. D. Solares, D. B. Newell, N. B. Zhitenev, and J. A. Stroscio,
Science {\bf 336} 1557 (2012). 

\bibitem{Tomori} 
H. Tomori, A. Kanda, H. Goto, Y. Ootuka, K. Tsukagoshi, S. Moriyama, E. Watanabe, and D. Tsuya, Appl. Phys. Express {\bf 4}, 075102 (2011). 

\bibitem{KBA2011} K.-J. Kim, Ya. M. Blanter, and K.-H. Ahn, Phys. Rev. B {\bf 84}, 081401(R) (2011).

\bibitem{Abedpour} N. Abedpour, R. Asgari, and F. Guinea, Phys. Rev. B
{\bf 84}, 115437 (2011). 

\bibitem{Wakker} G. M. M. Wakker, R. P. Tiwari, and M. Blaauboer, Phys. Rev. B {\bf 84}, 195427 (2011).

\bibitem{Gonzalez} J. Gonz\'alez, F. Guinea, and  M. A. H. Vozmediano, Nucl. Phys. B {\bf 424}, 595 (1994).

\bibitem{Fer1} F. de Juan, M. Sturla, and M. A. H. Vozmediano, Phys. Rev. Lett. {\bf 108}, 227205 (2012).

\bibitem {Pereira} V. M. Pereira, A. H. Castro Neto, and N. M. R. Peres, Phys. Rev. B {\bf 80}, 045401 (2009).

\bibitem{Alex} A. L. Kitt, V. M. Pereira, A. K. Swan, and B. B. Goldberg, Phys. Rev. B {\bf 85}, 115432 (2012); see important erratum at Phys. Rev. B {\bf 87}, 159909(E) (2013).

\bibitem{Fer-condmat} 
F. de Juan, J. L. Ma\~nes, M. A. H. Vozmediano,  Phys. Rev. B {\bf 87}, 165131 (2013).

\bibitem{Masir} 
M. R. Masir, D. Moldovan, F. M. Peeters, arXiv:1304.0629v2 (2013).

\bibitem{Salvador} 
J. V. Sloan, A. A. P. Sanjuan, Z. Wang, C. Horvath, and S. Barraza-Lopez, Phys. Rev. B {\bf 87}, 155436 (2013).
 
\bibitem{Roy1} 
B. Roy, Phys. Rev. B {\bf 84}, 035458 (2011).

\bibitem{Roy2} 
B. Roy, Z. X. Hu, and K. Yang, Phys. Rev. B {\bf 87}, 121408(R) (2013).

\bibitem{Morpurgo} P. Recher, B. Trauzettel, A. Rycerz, Ya. M. Blanter, C. W. J. Beenakker, and A. F. Morpurgo, Phys. Rev. B {\bf 76}, 235404 (2007).

\bibitem{Cong} C.-H. Yan and L.-F. Wei, J. Phys.: Condens. Matter {\bf 22},
295503 (2010).

\bibitem{Richter} 
J. Wurm, M. Wimmer, H. U. Baranger, and K. Richter, Semicond. Sci. Technol. {\bf 25}, 034003 (2010).

\bibitem{Russo} 
S. Russo, J. B. Oostinga, D. Wehenkel, H. B. Heersche, S. S. Sobhani,
L. M. K. Vandersypen, and A. F. Morpurgo, Phys. Rev. B {\bf 77}, 085413 (2008).

\bibitem{Ihn1} M. Huefner, F. Molitor, A. Jacobsen, A. Pioda, K. Ensslin, and T. Ihn, Phys. Status Solidi B {\bf 246}, 2756 (2009).

\bibitem{Ihn2}M. Huefner, F. Molitor, A. Jacobsen, A. Pioda, C. Stampfer, K. Ensslin, T. Ihn, New J. Phys. {\bf 12}, 043054 (2010).

\bibitem{Haug} D. Smirnov, H. Schmidt, and R. J. Haug,
Appl. Phys. Lett. {\bf 100}, 203114 (2012).

\bibitem{Rahman} A. Rahman, J. W. Guikema, S. H. Lee, and N. Markovi\'c, 
Phys. Rev. B {\bf 87}, 081401(R) (2013).

\bibitem{Trauzettel} 
J. Schelter, P. Recher, B. Trauzettel, Solid State Commun. {\bf 152}, 1411 (2012).

\bibitem{Berry} M. V. Berry and R. J. Mondragon, Proc. R. Soc. A 
{\bf 412}, 53 (1987).

\bibitem{Beenakker1} C. W. J. Beenakker, Rev. Mod. Phys. {\bf 80}, 1337 (2008).

\bibitem{Baranger}
 J. Wurm, A. Rycerz, \.{I}. Adagideli,
M. Wimmer, K. Richter, and H. U. Baranger, Phys. Rev. Lett. {\bf 102}, 056806 (2009).

\bibitem{Peres} N. M. R. Peres, J. N. B. Rodrigues, T. Stauber, and J. M. B.
Lopes dos Santos, J. Phys. Cond. Matt. {\bf 21}, 344202 (2009).

\bibitem {Ando} H. Suzuura and T. Ando, Phys. Rev. B {\bf 65}, 235412 (2002).

\bibitem {Landau} L. D. Landau and E. M. Lifshitz, {\it Theory of Elasticity}
(Pergamon Press, Oxford, 1970).

\bibitem{Fer2} F. de Juan, A. Cortijo, and M. A. H. Vozmediano, Phys. Rev. B {\bf 76}, 165409 (2007).

\bibitem{Zarenia2} 
M. Grujic, M. Zarenia, A. Chaves, M. Tadic, G. A. Farias,  and F. M. Peeters, Phys. Rev. B {\bf 84}, 205441 (2011).


\end{thebibliography}
\end{document}